# Magneto-elastic Polarons and Superconductivity of Underdoped Cuprates HTS's

G.G.Sergeeva

National Science Center "Kharkov Institute of Physics and Technology", Kharkov, Ukraine

**Abstract**

It is discussed some problems of the normal state of high temperature cuprates superconductors (HTS's) with $d$-wave pairing and next suppositions: 1) at pseudo-gap temperature $T^*$ it occurs the dimensional crossover from 3D incoherent normal state to quasi 2D system of $CuO_2$ layers; 2) "naked holes" are bad quasiparticles for quasi 2D system with Jahn-Teller and mixed-valent $Cu$ ions, strong electron correlations and fluctuations, and inherent strong $p-d$ hybridization of copper and oxygen orbitals; 3) this leads to reducing of "naked holes" to two types of magneto-elastic polarons. It is shown that one of them, ferromagnetic bound polarons, lead to the formation of the stripes at $T < T^*$, and the attraction between antiferromagnetic polarons lead to their pairing and to superconducting fluctuations in $CuO_2$ plane. Some experimental evidences of the observation of ferromagnetic bound polarons and of the antiferromagnetic bipolarons are discussed.

## Introduction

Crucial role of $d$-electrons in the unusual properties of transition metal compounds at least fifty years is intensively discussed in condensed state physics. The interest to them is caused by two important problems with studying of the nature of colossal magneto-resistance materials, such as manganites; and the nature of high temperature superconductivity in general, and essentially for cuprates HTS's with d-wave pairing. Nearly 15 years high temperature superconductivity still remains not being understood theoretically. And now even new questions are addressing to the normal state of HTS's:

i) what is the nature of "pseudogap" state at $T^* \gg T_c$?

ii) what is the "stripe states"?

This paper is an attempt for underdoped HTS to discuss these problems and next suppositions: at temperature $T^*$ it occurs the dimensional crossover of normal state from 3D incoherent state to quasi 2D system of $CuO_2$ layers; for such system with Jahn-Teller and mixed-valent $Cu^{+2}$ and $Cu^{+3}$ ions strong electron correlations and fluctuations, and inherent strong $p-d$ hybridization of copper and oxygen orbitals are making bad quasiparticles out of "naked holes" and lead to reducing their to magneto-elastic polarons.

## Dual character of $p-d$ hybridization in perovskites oxides of transition metals

First of all it is need to understand why the same d-electrons in $Sr_2RuO_{4.1}$ lead to p-wave pairing, and in cuprate HTS's, for example in $La_2CuO_{4+x}$, lead to d-wave pairing. The cause of this difference is lying in the crucial role of $p-d$ hybridization for these compounds. It is known that oxides of transition metal (TM) have inherent strong $p-d$ hybridization of TM and oxygen orbitals with dual character of results: $\pi$-bonds with $(d_{xy}-p)$ hybridization lead to direct ferromagnetic (FM) exchange ($I<0$) and p-wave superconductivity, and $\sigma$-bonds with $(d_{x^2-y^2})$ hybridization lead to indirect anti-ferromagnetic (AFM) exchange ($J>0$), and d-wave superconductivity. Symmetry of order parameter depends on the character of the states near Fermi energy $E_F$: for cuprates they have $(d_{x^2-y^2}-p)$ character, and $J>>I$, that leads to d-wave superconductivity; but for ruthenites they have $(d_{xy}-p)$ character, and $I>>J$, that leads to p-wave superconductivity. In simple $t-J-I$ model of $p-d$ hybridization [1] with AFM and FM interactions it was shown that temperature of superconducting transition with p-wave pairing, $T_c(p)$, is more less than temperature with d-wave pairing $T_c(p)<<T_c(d)$ at the equal interaction constants.

The main problem in normal state of HTS is its incoherent behavior, which point out that "naked holes" are bad quasiparticles because they are strong interacting with localized spins of copper and moving oxygen ions. At first L.D.Landau [2] pointed out that even in ideal crystal strong interactions between quasiparticles and moving ions lead to new type of quasiparticles, which later were named polarons. Magnetic polarons at first were introduced by E.L.Nagaev in 1967 (see refs. in review [3]). It is known that for perovskite copper oxides there are two type polarons: inherent strong $p-d$ hybridization of ions orbitals leads to AFM Zhang-Rice polarons (ZRP's) [4], and FM polarons (FMP's) [3]. Intermediate-size FMP and phase separation in the copper oxides were investigated in ref.[5]. Existence of localized polaronic states depends on the dimensionality of system and on value of the amplitude of $p-d$ hybridization. It is known that charge transfer along $c$-axis for underdoped HTS has incoherent character and this is a result of thermal fluctuations at

$$k_B T > t_c^2(T)/t_{ab} \qquad (1)$$

Here $t_c$ and $t_{ab}$ are the strength of interlayer and intralayer couplings of the charges, $k_B$ is Boltzman constant. At the temperature decreasing thermal fluctuations limit out the interlayer tunneling, and at

$$k_B T \approx t_c^2(T)/t_{ab} \qquad (2)$$

this leads to dimensional crossover when the charge becomes "two-dimensional" at the temperature

$$T^* = t_c^2(T^*)/k_B t_{ab} \qquad (3)$$

At $T < T^*$ the change of the metallic character of the resistivity along $c$-axis on semiconductive one occurs. This is very important conclusion because for 2D systems any localized states (including polaronic states) exist at any value of the interaction constants.

## The Charge Hamiltonian in $CuO_2$ plane

We consider a Hamiltonian describing a charge in single $CuO_2$ layer

$$H_{ef} = H_{DE} + H_{JT} \qquad (4)$$

where

$$H_{DE} = \frac{1}{2}\sum J_{n+\rho}\vec{S}_n\vec{S}_{n+\rho} - J_H \sum \vec{S}_n a^+_{n,\lambda,\sigma}\hat{\sigma} a_{n,\lambda,\sigma} - \sum t^{\lambda\lambda'}_{n,n+\rho} a^+_{n,\lambda,\sigma} a_{\rho,\lambda'\sigma} \qquad (5)$$

Here is phenomenological Kondo Hamiltonian [6] with superexchange interactions $J_{n,n+\rho}$ of spins $\vec{S}_n$ neighboring $Cu$ ions with oxygen ions through $p_\sigma$ bonds, which has AFM character, $J_{n,n+\rho} > 0$, and through $p_\pi$ bonds with FM character ($J_{n,n+\rho} \equiv I_{n,n+\rho}$). The former AFM interaction leads to effective integral of charge transfer $t^{\lambda\lambda'}_{n,n+\rho} = t_{AFM}$. Here $t_{AFM}$ is hybridization amplitude, which is proportional to overlapping of orbitals wave functions ($d_{x^2-y^2} - p$). FM interaction leads to $t^{\lambda\lambda'}_{n,n+\rho} = t_{FM}$, where $t_{FM}$ is proportional to overlapping of orbitals wave functions ($d_{xy} - p_\sigma$). In (5) second term $J_H$ is Hund exchange, $a_{n,\lambda,\sigma}$ are the hole operators, corresponding to hole at atom with position $n$ and spin $\sigma$ in the state $\lambda$, and third term is tunnel charge transfer between $Cu$ ions.

Second component in (4) is the phenomenological Hamiltonian of the charge interaction with Jahn-Teller distortion of oxygen ions:

$$H_{JT} = g_{JT}\sum[a^+_{n,\lambda,\sigma} Q^{\lambda\lambda'}_n(j) a_{n,\lambda',\sigma} + \frac{k_{JT}}{2}Q_n^2(j) + \frac{M_{JT}}{2}(\frac{dQ_n(j)}{dt})^2] \qquad (6)$$

Here $g_{JT}$ is the constant of elastic Jahn-Teller (JT) interaction, $Q_n(j)$ is the operator of $j$ normal JT mode, $k_{JT}$ and $M_{JT}$ are their elastic constant and effective mass. Canonical Holstein-Lang-Firsov transformation [7] let us in (6) to get of first term, and leads to new stationary states which are magneto-elastic Jahn-Teller polarons with renormalizated integral of charge transfer

$$t \to t_{JT} = t \exp[-\frac{g_{JT}^2}{\Omega_{JT}^2}(1 + 2n_{JT})] \qquad (7)$$

Here $n_{JT}$ is mean phonon number with frequency $\Omega_{JT} = \sqrt[2]{k_{JT}/M_{JT}}$.

## Jahn-Teller renormalization of Zhang-Rice polarons

Superexchange interaction $J_{n,n+\rho}$ of spins $\vec{S}_n$ of neighboring $Cu$ ions with oxygen ions through $p_\sigma$ bonds ($d_{x^2-y^2} - p$) has AFM character and is equal [4]

$$J_{n,n+\rho}^{AFM} = 4\overline{t_{AFM}}^4 \varepsilon_p^{-2}(1/U + 1/2\varepsilon_p) > 0 \qquad (8)$$

Here $\overline{t_{AFM}}$ is effective integral of charge transfer $t_{AFM}$ with taking into account the renormalization (7); $\varepsilon_p > 0$ is the atomic energy of oxygen hole; and $U$ is the on-site Coulomb repulsion at $Cu$ site [8]. As it was shown by F.C.Zhang and T.M.Rice [4] $p-d$ hybridization strongly binds a hole on each square of $O^2$ ions to the central $Cu^{+2}$ ion to form a local singlet, ZRP. Taking into account the renormalization (8) of the interaction $J_{n,n+\rho}^{AFM}$ let us below to name this local singlet as Jahn-Teller ZRP magneto-elastic polaron (JT ZRP). This polaron has energy $\simeq -15,68 t_{ef}$ and moves through the lattice with effective nearest-neighbor hopping $t_{ef} \sim \overline{t_{AFM}^2} \varepsilon_p^{-1}$ which depends on renormalized integral of charge transfer.

The depth of JT ZRP level is exceeded $d-d$ exchange energy $E_{dd} = 4JS^2$. JT ZRP has the important advantage over "naked hole", because its state is phase coherent state; therefore JT ZRP's are really good quasiparticles with large value of coupling with phonons. The reason of large coupling is obtained at small holes concentration $n < n_c$ where $n_c \sim 10^{20} cm^{-1}$. It is known that at this condition plasma edge lies below the highest optical phonons and they are unshielded.

## Jahn-Teller renormalization of bound ferrons

The studying of FM self-trapped states of a charge carrier in a doped AFM crystal was began by E.M.Nagaev which in 1968 proposed the models of free FMP and bound ferron (see refs. in [3]). For free ferron in layered AFM crystal strong magnetic anisotropy along c-axis leads to the absence of well defined FM region of a finite size inside of which the d-electron is localized. In the frame of I.M.Lifshits theory for 2D system discrete level exists at any values of $t$ and $I$ [9]. But it is too shallow to compensate the loss in the $d-d$ exchange energy, because the quantity of the ratio

$$|I/J| \sim J_H/U \leqslant 0.2 \tag{9}$$

is small and leads to the depth of the level which is insufficient for this compensation. Therefore for doped layered anisotropic antiferromagnets the bound ferrons (BFMP) consideration is more accurate as it was shown in ref.[3]. The BFMP is corresponding to a renormalized hole which is localized at an impurity neighborhood, e.g. it is a usual quasi-local state for disordered state [10]. Creation of a magnetized region around the impurity additionally diminishes the energy of the system. For cuprate HTS's mixed valent $Cu^{+n}$ ions with $n \neq 2$ can serve as the localization centers of BFMP's.

In Ref. [3] it was shown that electron wave function $\Psi(r)$ in magnetized region should be determined from the same equation as for the electron in a hydrogen atom with an additional requirement $\Psi(r) = 0$ at $r = R$, where $R$ is the ferron radius or the size of FM microregion:

$$(-\frac{\Delta}{2m} - \frac{e^2}{\varepsilon r} - U - E_s)\Psi(r) = 0 \tag{10}$$

Here $E_s$ is the energy of the bound state, $m = 1/\overline{t_{FM}}a^2$ is the electron effective mass, $\varepsilon$ is the dielectric constant, $a$ is $Cu-O$ distance in $CuO_2$ plane, and $U = 2\overline{t_{FM}}$ is the spherical potential well, where $\overline{t_{FM}}$ is effective integral of charge transfer $t_{FM}$ with taking into account the renormalization (7). The ground state wave function $\Psi(r)$ of Eq.(10) is given in Ref.[11], and expresses through the confluent hyper-geometric Kummer function $\Phi(1-n, 2, \frac{2r}{na_B})$[12] where $n$ is argument of this function, and $a_B$ is Bohr radius. Analysis of these functions shown, that the minimal ferron energy is lower than the energy of the bound s electron, and the size of the bound ferron is equal $R \cong 2\varepsilon\overline{t_{FM}}(a/e)^2$. With taking into account value $\varepsilon = 9$ of the dielectric constant for $CuO_2$, we can estimate value $R \sim 2\varepsilon a_B \geqslant 10\text{Å}$.

## Discussion

For J.G.Bednorz and K.A.Muller the idea that the Jahn-Teller polarons with strong electron-phonon coupling might be important for high temperature superconductivity was the original direction of their search [13]. For TM oxydes, and ceramics $BaBi_xPb_{1-x}O_3$ the existence of the attraction interaction between the polarons and their pairing with the formation of bipolarons were firmly determined 20 years ago (for example, see refs. in review [14]) [15]. It is important to note that for TM oxides the observation of charge ordered state was bound up with bipolarons [15]. At first the conditions of the transition of systems with bipolarons to the superconducting state at the exchange by optical phonons was discussed by L.N.Bulaevskii [16], which shown that for this transition the frequencies of optical phonons, and the energy of electron-phonon coupling, and Coulomb energy must be the values $\sim 0.1$ ev. For HTS's the polaronic pairing are intensively discussed (see refs. in review [17]).

Above it is shown that dual character of inherent strong $p-d$ hybridization in perovskites of transition metals leads to two types of polarons. One of them is mobile AFM Jahn-Teller-Zhang-Rice polaron, and second is bound FM polaron (or Nagaev bound ferron). Below it will be shown that both types of polarons are very important for HTS's.

1. Stripes generation for underdoped HTS.

At $T < T^*$ charge becomes "two-dimensional", and strong Jahn-Teller distortion, and $(d_{xy} - p)$ hybridization lead to direct FM exchange orbitals of $d$-electrons of neighbor $Cu$ ions and to the generation of bound ferrons in $CuO_2$ plane. If the size of bound ferron is compared with mean distance $R \sim ax^{-1/3}$ between neighboring localization centers, $Cu^{+n}$ ions, ($x$ is the doping concentration), chains of BFMP's make up in $CuO_2$ planes the net of stripes along which only pairing polarons, e.g. bipolarons, can move because JT ZR polarons destroy potential well which was created by BFMP. For underdoped HTS's with dopant concentration $x \leqslant x_{cr}$ this leads to the generation of the stripe structure in $CuO_2$ planes. With using of the estimation of the size of the bound ferron $R$ in above section

$$x_{cr} \sim (a/R)^3 \sim (\frac{e^2}{2a\varepsilon \bar{t}_{FM}})^3 \qquad (11)$$

2. JT ZR polarons pairing as a mechanism of high temperature superconductivity.

Attraction between the JT ZR mobile polarons at $T < T^*$ can lead to the bipolarons formation in $CuO_2$ plane and to two dimensional superconducting fluctuations, at Bulaevskii conditions [16] that the frequencies of Jahn-Teller cooperative oxygen displacements, and the energy of electron-phonon coupling, and Coulomb energy are the values of the same order. These fluctuations lead to the decreasing of tunneling probability of charge along $c$-axis, $t_c(T)$, with the temperature decreasing [18]. At sufficiently small $t_c(T)$ inequality $T_c/E_F \geqslant t_c(T_c)$ determines the temperature of superconducting transition which occurs according to the Kats scenario for the layered system [19]. Thus, the attraction of mobile JT ZR polarons can leads to the high temperature superconductivity, and all above enumerated interactions together are important participants of this phenomena.

These circumstances make very difficult the purely analytical determination even of the true ground state for HTS's, and therefore the search of pairing mechanism is such prolonged and hard work. This is the reason why it is necessary for the determination of the true ground state to use the experimental possibilities.

3. About the possibility of the experimental observation of polarons and bipolarons for cuprates HTS's.

At first the experimental evidence of dual nature of electron structure was received with using high resolution ARPES spectra of $La_{2-x-y}Nd_ySr_xCuO_4$ and $La_{2-x}Sr_xCuO_4$ [20]. The existence of spectral weight of localized states along the nodal direction where superconducting gap is zero in $d$-wave symmetry of the itinerant states was shown. The unique known type of quasiparticles with dual nature are polarons which are localized in finite region and can freely move as the itinerant states. Second important the experimental evidence of large value of the coupling polarons with phonons was received also from ARPES spectra of $Bi_2Sr_2CaCuO_8$, $Bi_2Sr_2CuO_6$ and $La_{2-x}Sr_xCuO_4$ [21] where it was determined the phonons frequencies $\sim 0.05 - 0.08 eV$.

It was shown that polaronic states can be identified from optical absorption data in ref. [5], where at first an intermediate-size of FM polaron was identified after the evaluation of infrared absorption spectra with using of the adiabatic random phase approximation. Bipolaronic states introduce the split of absorption spectra of polaron states. Recently [22] at the studying absorption spectra of monocrystilline YBCO films an additional maximum on high-energy side of AFM main maximum was observed at $T < T^*$. It was quite naturally to suppose that observation of charge ordered state are bound up with pairing AFM polarons [15, 23]. The difference between the energy of both split maxima is equal 0.14 eV that evidences about their magnetic coupling. The magneto-elastic nature of the bipolarons is confirmed by the observations [21,22] from which one can see that the energy of the coupling polarons with phonons is the same order as the magnetic coupling and both approximately equals 0.1 eV, as it was required by Bulaevskii conditions [16]. It is interesting to note that at the further temperature decreasing this split does not change up to $T = T_c$ when the intensity of both split maxima sufficiently increases. The latter was due by the transition from the state with 2D superconducting fluctuations in $CuO_2$ planes to the three dimensional coherent superconducting states [18,19] that leads to the increase of the bipolarons number.

REFERENCES

1. E.V.Kuz'min, S.G.Ovchinnikov, I.O.Baklanov. Zh.Eksp.Teor.Fiz., 1999, **116** 65.

2. L.D.Landau. Sow.Phys. 1933, **33** 664.

3. E.L.Nagaev. Phys.Rev. 1999, **B60** 455.

4. F.C.Zhang and T.M.Rice. Phys.Rev.1988, **B37** 3759.

5. K.Yonemitsu and A.R.Bishop, J.Lorenzana. Phys.Rev.Lett. 1992, **69** 965.

6. A.Ramsak and P.Prelovshek. Phys.Rev. 1990, **B 42** 10415.

7. I.G.Lang, Yu.A.Firsov. Zh.Eksp.Teor.Fiz. 1962, **43** p.1843.

8. P.W.Anderson. Phys.Rev. 1959, **115** 2.

9. I.M.Lifshits. Usp.Fiz.Nauk, 1964, **83** 617.

10. G.G. Sergeeva. Zh.Eksp.Teor.Fiz. 1965, **48** 158.

11. L.D.Landau and E.M.Lifshits. Kvantovaya Mekhanika (Quantun Mechanics), Nauka, Moscow, 1989, 767p.

12. E.Janke, F.Emde, and F.Loesch. Edited by B.Teubner (Ver-laggesellschaft, Stuttgard, 1960).

13. J.G.Bednorz and K.A.Muller. Science, 1987, **237** 1133.

14. M.I.Klinger. Usp.Fiz.Nauk, 1985, **146** 105.

15. B.K.Chakraverty. Phil.Mag.B, 1980, **42** 473.

16. L.N.Bulaevskii et.al. Zh.Eksp.Teor.Fiz., 1984, **87** 1490.

17. A.S.Alexandrov, A.B.Krebs. Usp.Fiz.Nauk, 1992, **162** 1.

18. G.G.Sergeeva et al. Low Temp.Phys. 2001, **27** 634.

19. E.I.Kats, Sov.Phys.JETP, 1969, **29** 896

20. X.J.Show, T.Yoshida, S.A.Keller et al. Phys.Rev.Lett., 2001, **86** 5578.

21. A.Lanzara et al.Nature 2001, **412** 510.

22. V.V.Eremenko, V.N.Samovarov et al. Low Temp.Phys. 2001, **27** 1327.

23. V.V.Eremenko et al. Low Temp.Phys. 2002, **28** (to be published).